# Hydrogen storage in C14 type $Ti_{0.24}V_{0.17}Zr_{0.17}Mn_{0.17}Co_{0.17}Fe_{0.08}$ high entropy alloy

Abhishek Kumar[1], T. P. Yadav[1,2*], M.A. Shaz[1] and N.K. Mukhopadhyay[3]

[1]Hydrogen Energy Centre, Department of Physics, Institute of Science
Banaras Hindu University, Varanasi, Uttar Pradesh, India
[2]Department of Physics, Faculty of Science, University of Allahabad, Prayagraj-211002, India
[3]Department of Metallurgical Engineering, Indian Institute of Technology (Banaras Hindu University),
Varanasi-221 005, India

## Abstract

In this present investigation, we discussed the synthesis, microstructure, and hydrogen storage behavior in C14 type intermetallic Laves phase in a hexanary $Ti_{0.24}V_{0.17}Zr_{0.17}Mn_{0.17}Co_{0.17}Fe_{0.08}$ high entropy alloy (HEA). In this HEA, three elements are hydride-forming elements (Ti, V, Zr), whereas other three are non-hydride-forming elements (Fe, Mn, Co). The thermodynamic parameter like enthalpy of mixing was calculated using the Meidma's model. The mixing enthalpy ($\Delta H_{mix}$) of $Ti_{0.24}V_{0.17}Zr_{0.17}Mn_{0.17}Co_{0.17}Fe_{0.08}$ HEA system was evaluated to be- 23.3472 kJ/mole, and atomic radius mismatch turned out to be = 7.441%. This alloy was synthesized using 35 kW radio frequency induction furnace under argon atmosphere. X-ray diffraction technique (XRD) revealed that this system belongs to the C14 type Laves phase with unit cell parameters a= b =5.0158 Å, c=8.1790 Å, $\alpha = \beta = 90°$, $\gamma = 120°$ under Space group $P6_3/mmc$. Microstructural analysis was carried out with the help of a transmission electron microscope (TEM). The SEM- EDX data confirms the elemental composition. Hydrogen absorption and desorption of this high entropy intermetallic was carried out using the PCI apparatus. The total hydrogen storage of this system was observed around ~0.53 wt%. However, it exhibited better hydrogen and ab/de-sorption kinetics. With the help of the Van't Hoff plot, calculated experimental change in enthalpy of $Ti_{0.24}$-$V_{0.17}$-$Zr_{0.17}$-$Co_{0.17}$-$Fe_{0.08}$-$Mn_{0.17}$ HEA for hydrogen absorption and desorption was found out to be ~ -19.06 ± 1.12 kJ/mol and -34.10 ± 1.32 kJ /mol respectively. The possibility of developing high entropy Laves phase-based hydrogen storage materials was advocated.

Corresponding authors: yadavtp@gmail.com

**Introduction**

Recent years have seen a lot of interest in a new class of materials called 'High Entropy Alloys' (HEAs) (Marques *et al.* 2021, Yadav *el al.* 2017, Mishra *et al.* 2019, Mishra *et al.* 2020). In general, HEAs contain five or more elements, each with a concentration of five to thirty-five atomic percentages (at.%) or more, in contrast to conventional alloys based on a single primary element. To improve phase stability, HEAs are understood to exhibit large mixing entropies of solid solution phases (Murty *et al.* 2019). The research publication by Yeh *et al.* (2004a 2004b), Cantor *et al.* (2004), and Ranganathan (2003) was published for the first time for launching the field of HEAs. Yeh independently proposed the single-phase multi-principal element alloy in 1995, making this idea a ground-breaking success in researching HEAs (Murty *et al.* 2019). It' is interesting to note that the high mixing entropy in multi-principal element alloys can dramatically lower the number of phases in high-order alloys, leading to a single phase solid solution (Tsai *et al.* 2014). HEA has many functional properties like magnetic,, thermoelectric, catalytic, hydrogen storage etc. In these functional properties, hydrogen storage is considered to be one of the interesting areas to explore the HEA as an effective hydrogen storage material. Nowadays, in order to counteract climate change and the rise in global warming brought on by conventional fossil fuels; people demand innovative, flexible, clean, and green energy sources. Among many fuels that are readily available worldwide, hydrogen is accepted as one of the best candidates due to its high energy range per unit mass. Three essential elements that are needed to use hydrogen as a fuel in the future are (i) hydrogen production, (ii) its storage, and (iii) applications. Hydrogen storage is one of the most crucial components of using hydrogen as a fuel. One of the safest and most efficient ways to store hydrogen is in solid-state metal hydrides. Due to the infinite combination of alloy forming possibilities, the HEAs are novel and promising materials for hydrogen storage (Yadav *et al.* 2022). In 2010, the first investigation was done in HEAs to study the hydrogen storage kinetics. This study claimed 0.03-1.80 wt% hydrogen storage in multi-principal component $CoFeMnTi_xV_yZr_z$ (Kao *et al.* 2010) alloys; after that, in TiZrHfNbV HEA, 2.7wt% hydrogen storage was reported in 2016 (Sahlberg *et al.* 2016). There is only one BCC phase in this alloy composition. One more point common in this system is that this alloy system is designed with all the hydride forming elements, because of which it has a good hydrogen storage capacity. In recent years hydrogen storage is reported as high as 3.51 wt% in $V_{35}Ti_{30}Cr_{25}Fe_5Mn_5$ HEA belonging to a single BCC phase (Liu *et al.* 2021). On the contrary, the maximum hydrogen storage in Laves phases is known to be 1.91 wt% (Sarc *et al.* 2020). It can stated from the reported data that the Laves phase has less storage properties and better absorption and desorption kinetics compared to BCC phase. The investigation on low-vanadium TiZrMnCrV-based alloys for high-density hydrogen storage (Zhou et al. 2021) was reported. Due to its maximal interstitial sites available for absorbing hydrogen in their voids, C14 Laves phase has been explored as hydrogen storage phase tested in recent study. People have recently been concentrating on the research of phase stability during hydrogen absorption and desorption of HEAs. In multi-component HEA for TiZrFeMnCrV (Chen et al. 2022), C14 type Laves phase-based HEA was fabricated and followed by hydrogen storage testing after mechanical milling. The maximal hydrogen absorption for this alloy was reported to be 1.80 wt% for the first cycle and 1.76 wt% for the second cycle. According to their findings, the hydrogen storage capacity varied marginally between each cycle's i.e., 1.76 and 1.73 wt%. In another study, TiZrCrMnFeNi HEA of C14 Laves phase has exhibited hydrogen absorption as 1.7 weight percent (Edalati et al. 2020). Kumar et al (2022) has shown that TiZrVCrNi Laves phase with 1-52 weight percent hydrogen remains stable even after 10 cycles of hydrogenation from the perspective of phase stability. The TiZrNbCrFe HEA consisting of C14 Laves phase as maor and BCC phase as minor was reported by Floriano et al. 2021 to have 1.9 wt% hydrogen storage capacity.In view of the potential of HEAs for hydrogen storage capability, it was felt worth pursuing the study of other high entropy based alloys for exploring their structure and hydrogen storage performance. Accordingly, in the present study, we selected TiZrVMnFeCo nonequiatomic HEAs and investigated the structure, microstructure, and hydrogen storage kinetics. We chose a HEA system with three hydride forming elements (TiZrV) and the remaining three non-hydride-forming elements (Mn, Fe, Co).The thermodynamic calculation for evaluating enthalpy of mixing of this HEA was done using Meidma model. This HEA was synthesized with the help of a 35 KW Radio Frequency Induction furnace in the argon atmosphere and characterized by XRD, SEM and TEM techniques Hydrogen storage performance was evaluated using pressure composition isotherm (PCI) equipment supplied by Advanced Material Corporation (Pittsburgh, USA).

**Material synthesis and characterization methods**

The high purity materials powder for the synthesis of the $Ti_{0.24}V_{0.17}Zr_{0.17}Mn_{0.17}Co_{0.17}Fe_{0.08}$ HEA system was procured from Alfa Aesar with a purity of more than 99.50%. The constituent elements were taken as per their stoichiometry for making a palette using a cylindrical steel mold equipped with the hydraulic press of acting pressure ~3x105 N/m$^2$. The palette (~10 g by weight) then used for the as-cast synthesis of multicomponent HEA using the RF induction melting process under argon atmosphere (purity of more than 99.90%). The ingots are melted four times to ensure uniformity of chemical composition. The as-cast induction melted ingots of HEA crushed and converted into powder form to perform further characterization. The first cutting-edge characterization technique used for phase analysis is the Empyrean x-ray diffraction (XRD) system (Malvern Panalytical) equipped with an area detector (256x256 pixels) equipped with a graphite monochromator and Cu radiation source (CuKa; = 1.5406, operating at 45 kV and 40 mA) in Bragg-Brentano geometry. The transmission electron microscope (TEM), TECNAI 20 G$^2$, was used to acquire the microstructures and selected area electron diffraction (SAED) pattern of the samples operating at 200 kV of accelerating voltage.EVO 18 scanning electron microscope at operating voltage of 25 kV (vacuum $10^{-5}$ torr) was used to investigate surface morphology and perform energy dispersive X-ray analysis (EDX) as well as colour mapping of elements in the as-prepared samples. All de/re-hydrogenation measurements were carried out with the aid of an automated two-channel volumetric sieverts apparatus (supplied by Advanced Materials Corporation Pittsburgh, USA). For hydrogen storage testing, we took the 500 mg sample of HEA and placed the sample in the reactor seized by quartz wool. Before performing hydrogen cycle testing, the powder HEA sample was activated at 400℃ under a hydrogen pressure of 1/0.1 MPa for hydrogenation/dehydrogenation. After activation, testing of the hydrogen absorption kinetics at 410 °C under 60 atm $H_2$ pressure was carried out.

**Results and Discussion**

The experimental XRD diffraction patterns of the as-cast $Ti_{0.24}V_{0.17}Zr_{0.17}Mn_{0.17}Co_{0.17}Fe_{0.08}$ HEA are shown in figure 2(a). The diffraction profile has been recorded for the gross structural analysis of the as-cast alloy sample by using the Empyrean x-ray diffraction (XRD; Malvern Panalytical) system. All the diffraction peaks (shown in the figure. 2(a)) are well fitted with the hexagonal C14 Laves phase structure parameters.The XRD pattern was well refined through Le Bail profile fitting using JANA 2006 software shown in the figure. 2(b). The refinement data validated the $Ti_{0.24}V_{0.17}Zr_{0.17}Mn_{0.17}Co_{0.17}Fe_{0.08}$ HEA system with unit cell parameters of a=b= 5.0141 Å, c= 8.1756 Å, and the unit cell volume 178.0 Å3 under the space group of $P6_3$/mmc. All the refine parameters are given below in **Table 1**

To validate the structure analysis of this XRD, we used another characterization technique by transmission electron microscopy (TEM) for analyzing the phase and microstructure of this $Ti_{0.24}V_{0.17}Zr_{0.17}Mn_{0.17}Co_{0.17}Fe_{0.08}$ HEA**.** The bright field TEM micrograph of as-synthesized HEA shown in figure 3(a) identifies no other phases other than Laves phase. The corresponding SAD pattern of this as cast HEA shown in figure 3(b) validates that this HEA system belongs to a C14 type hexagonal structure with a corresponding space group is $P6_3$/mmc.

**Surface morphology and elemental composition analysis**

Scanning electron microscopy (SEM) has been done for surface microstructure and confirming homogeneous element distribution. Figure4 (a) shows the SEM –BSE, and Energy dispersive X-ray analyses (EDX) mapping images of as cast $Ti_{0.24}V_{0.17}Zr_{0.17}Mn_{0.17}Co_{0.17}Fe_{0.08}$ HEA with the corresponding region which is located in square box in figure 4(a). The SEM-BSE image reveals the microstructure of this HEA without any cracks or defects in this as-cast HEA. Figure 4(b) overlays all the constituent elements present in this HEA. EDAX mapping image establishes that all the constituent elements are distributed as per atomic percent in this as-cast $Ti_{0.24}$-$V_{0.17}$-$Zr_{0.17}$-$Co_{0.17}$-$Fe_{0.08}$-$Mn_{0.17}$ HEA. Figure 4(c) shows the SEM-BSE image from another region for the HEA sample, where no crack is observed, and also no other contrast corresponding another phase. Figure 4(d) shows the EDX elemental spectra to confirm the stoichiometry of the elements present in this as-cast HEA. All the data indicate that this HEA has forms a single Laves phase with uniform elemental distribution.

**Hydrogen ab/de-sorption analysis**

Hydrogen ab/de-sorption performance in as-cast $Ti_{0.24}V_{0.17}Zr_{0.17}Mn_{0.17}Co_{0.17}Fe_{0.08}$ HEA is studied in this section. The measurements of hydrogen sorption were carried out with automated two-channel volumetric sieverts instrument. The results of the absorption kinetic curve of the as-cast $Ti_{0.24}V_{0.17}Zr_{0.17}Mn_{0.17}Co_{0.17}Fe_{0.08}$ HEA are shown in figure 5(a). Before introducing hydrogen into as-cast HEA, we firstly activate the as-cast HEA under 400 ˚C under $10^{-3}$ atm evacuation. We perform hydrogenation at 410˚C under 60 atm hydrogen pressures. The hydrogen desorption kinetic curve of the as-cast Ti0.24-HEAis shown in figure 5(b). The hydrogen desorption kinetic curve of this as-cast HEA shows that this as-cast HEA absorbed 0.53 wt% of hydrogen within 15 seconds this curve. In contrast, the maximum storage capacity is evaluated to be about 0.72 wt% in 150 minutes. This fastest kinetics gives interesting results to understand the hydrogen storage performance In the case of desorption, we can see that the dehydrogenated curve shown in figure 5(b) the as cast $Ti_{0.24}V_{0.17}Zr_{0.17}Mn_{0.17}Co_{0.17}Fe_{0.08}$ HEA perform desorption at 410 ˚C under 1 atm hydrogen pressure. According to the hydrogenation desorption curve we can say that this HEA released 0.28 wt% hydrogen within one minute at 410 ˚C under 1 atm hydrogen pressure. The results suggests that this HEA shows faster hydrogen ab/desorption kinetics than some other Laves phase based HEAs.

The representative PCI ab/de-sorption of $Ti_{0.24}V_{0.17}Zr_{0.17}Mn_{0.17}Co_{0.17}Fe_{0.08}$ HEA has been shown in figure 6(a) the corresponding represents active Van't Hoff plots (shown in figure 6(b)). PCI was performed at 395˚C, 410˚C and 425˚C temperatures under 60 atm hydrogen pressures. With the help of the three different temperatures, we get the plot corresponding to temperature v/s pressure. Calculations of the entropy and enthalpy changes that occur throughout the hydrogen ab/de-sorption process typically employ the pressure values of the hydrogen ab/de-sorption platform at various temperatures. The change in enthalpy (ΔH) of hydride formation is given by the well-known Van't Hoff equation (Dornheim et al. 2010)

$$\ln P = \frac{\Delta H}{RT} - \frac{\Delta S}{R} \quad \ldots\ldots\ldots\ldots(i)$$

Where P is the previously specified plateau pressure, T is the corresponding temperature, R is the gas constant, and H and S are the reaction enthalpy and entropy changes, respectively. The alloys' Van't Hoff plots are computed using the P, as shown in figure 6. (b). The relationship between ln(P) and 1000/T is clearly linear, as can be seen in the image. The slope of the fitted curves for ln(P) and 1000/T as well as the intercept on the vertical coordinate allow for the quick calculation of the H and S. The results of the calculations demonstrate that the enthalpy of hydrogen desorption changes. The change in enthalpy of $Ti_{0.24}V_{0.17}Zr_{0.17}Mn_{0.17}Co_{0.17}Fe_{0.08}$ HEA for hydrogen absorption and desorption has been calculated to be $\Delta H_{abs}$~ -19.06 ± 1.12 kJ/mol and $\Delta H_{des}$ -34.10 ± 1.32 kJ/mol respectively. The smaller negative enthalpy of mixing in HEA suggests that they are more likely to form stable metal hydrides. The formation of the metal hydride's absorption and desorption enthalpies are not equal in the current experiment. Therefore, this system has fewer tendencies to create metal hydride and aids in improving the ab/desorption kinetics. This suggests that they have a decreased tendency to form a stable metal hydride.

**Conclusions**

In this study, we have successfully synthesized the hexanary $Ti_{0.24}V_{0.17}Zr_{0.17}Mn_{0.17}Co_{0.17}Fe_{0.08}$ HEA with the help of an RF induction furnace for the study of hydrogen storage kinetics. The evolution of a single phase of hexagonal C14 high entropy Laves phase with lattice parameters a = 5.01Å and c =8.17Å was established following Rietveld refinement in this multicomponent alloy. On the basis of the kinetics study, $Ti_{0.24}V_{0.17}Zr_{0.17}Mn_{0.17}Co_{0.17}Fe_{0.08}$ shows good ab/de-desorption kinetics (absorb ~ 0.53 wt.% of $H_2$ within 15 seconds) but poor in hydrogen storage capacity. The change in enthalpy of $Ti_{0.24}V_{0.17}Zr_{0.17}Mn_{0.17}Co_{0.17}Fe_{0.08}$ HEA for hydrogen absorption and desorption has been calculated to be ~ -19.06 ± 1.12 kJ/mol and -34.10 ± 1.32 kJ /mol respectively. The present investigation suggests the scope for further study on the hydrogenation kinetics at various temperatures for exploring the potential for developing Laves phase high entropy alloy for hydrogen storage.

**Acknowledgment**

The author (AK) wishes to thank the Council of Scientific and Industrial Research (CSIR) in New Delhi, India, for financial support for a senior research fellowship (Award No. 09/013(0952)/2020-EMR-I).

**Author contributions**

A.K. synthesized the materials and made the characterizations; T.P.Y. conceived, designed the experiments, organized the data and supervision. M.A.S. advised on the discussion of results; N.K.M. advised on the discussion of results and editing the manuscript. The manuscript was written through contributions of all authors. All authors have given approval to the final version of the manuscript.

**Notes**

The authors declare no competing financial interests.

**Figure captions**

**Figure 1**: (a) Schematic diagramof the synthesis protocol for $Ti_{0.24}V_{0.17}Zr_{0.17}Mn_{0.17}Co_{0.17}Fe_{0.08}$ HEA

**Figure 2:** (a) XRD pattern of $Ti_{0.24}$-$V_{0.17}$-$Zr_{0.17}$-$Co_{0.17}$-$Fe_{0.08}$-$Mn_{0.17}$ HEA system and (b) Rietveld refinement profile pattern of all the peaks well fitted with C14 type hexagonal parameters with unit cell parameters a= b =5.0158 Å, c=8.1790 Å, α = β = 90˚, γ = 120˚ under space group $P6_3/mmc$

**Figure 3 :** (a) TEM bright field micrograph of as-cast HEA synthesized by RF induction melting (b) Corresponding SAD patterns are shown indexed with hexagonal structure parameter under the space group of $P6_3/mmc$

**Figure 4:** (a) SEM–BSE and energy dispersive X-ray analyses (EDX) mapping images of as $Ti_{0.24}V_{0.17}Zr_{0.17}Mn_{0.17}Co_{0.17}Fe_{0.08}$ HEA (b) overlays all the constituent elements present in this HEA. (c SEM-BSE image from another region for the HEA. (d) EDX elemental spectra to validate the atomic percentage of the elements in this HEA.

**Figure 5:** (a) Hydrogenation curve of $Ti_{0.24}V_{0.17}Zr_{0.17}Mn_{0.17}Co_{0.17}Fe_{0.08}$ HEA at 410 ˚C under 60 atm $H_2$ pressure and (b) Dehydrogenation curve of hydrogenated $Ti_{0.24}V_{0.17}Zr_{0.17}Mn_{0.17}Co_{0.17}Fe_{0.08}$ HEA at 410 ˚C under 60 atm $H_2$ pressure

**Figure 6:** (a) Fig: (a) PCI ab/de-sorption curves of $Ti_{0.24}V_{0.17}Zr_{0.17}Mn_{0.17}Co_{0.17}Fe_{0.08}$ HEA and (b) Corresponding Van't Hoff plots for PCI ab/de-sorption curves.

**Table 1:**

Lattice Parameters and refinement parameters obtained from powder x-ray diffraction data of the as-cast HEA.

| | Refined Parameter and phase data |
|---|---|
| **Unit-Cell Parameters** | a= b =5.0158 Å, c=8.1790 Å, α = β = 90°, γ = 120° |
| **Space Group** | P6$_3$/mmc (Space Group = 194) |
| **R- Factor** | Rp = 3.23%, wRp = 4.45%, GOF = 1.26%, |
| **Volume** | V = 178.20Å$^3$ |

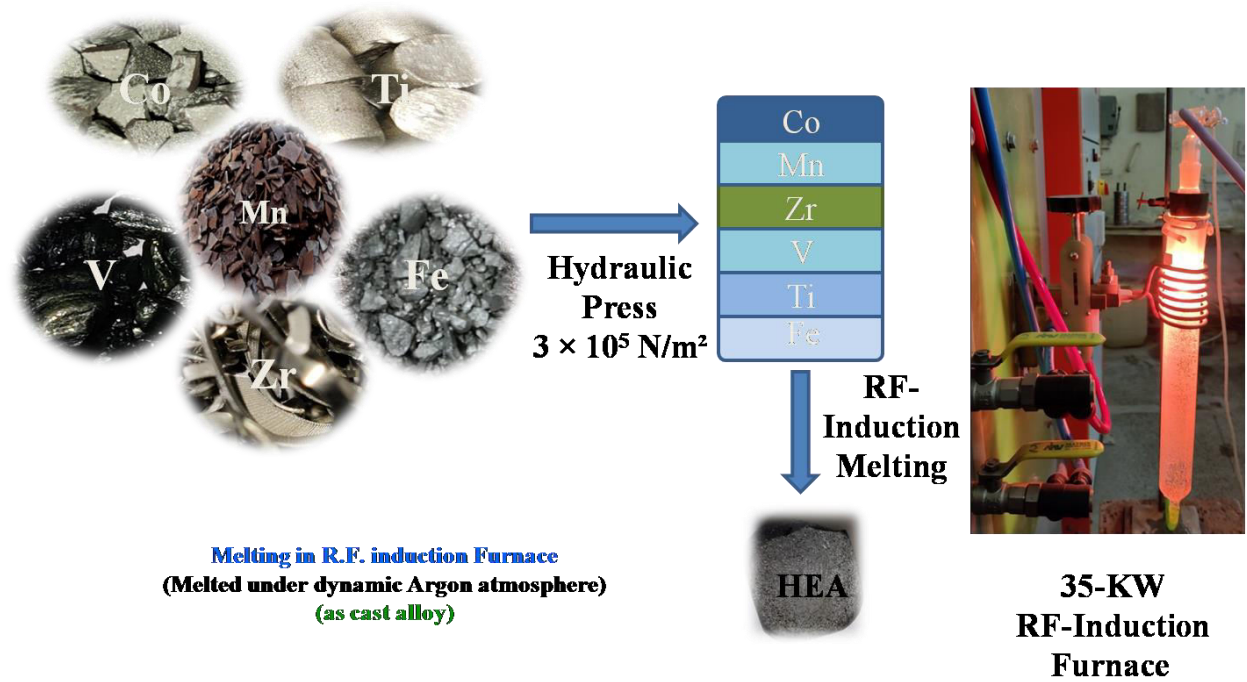

**Figure 1**: (a) Schematic diagram of the synthesis protocol for $Ti_{0.24}V_{0.17}Zr_{0.17}Mn_{0.17}Co_{0.17}Fe_{0.08}$ HEA

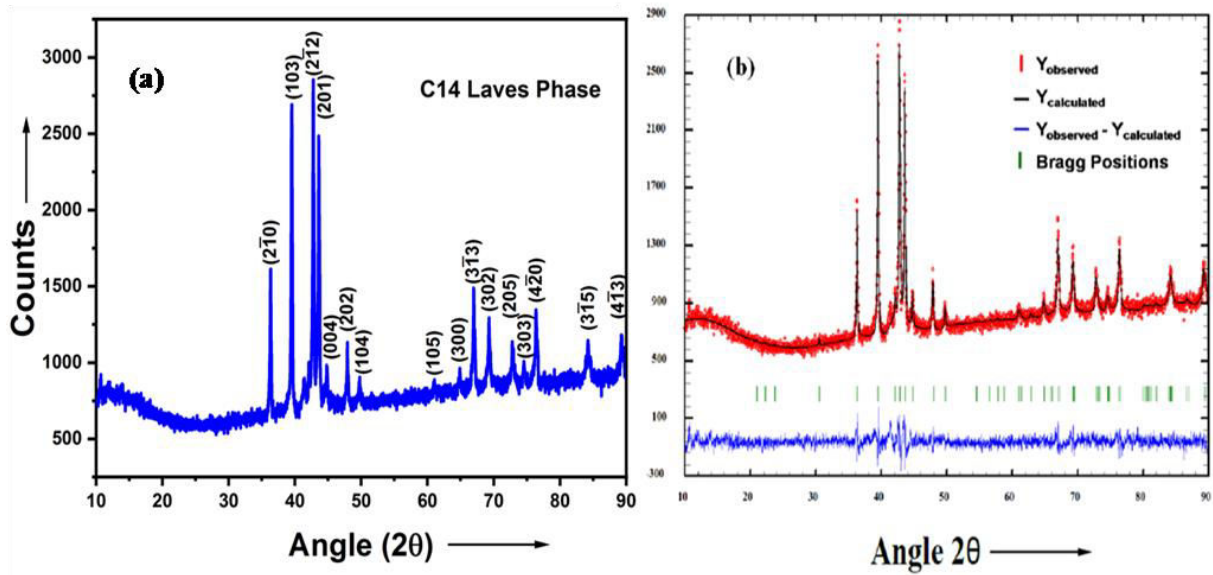

**Figure 2:** (a) XRD pattern of $Ti_{0.24}V_{0.17}Zr_{0.17}Mn_{0.17}Co_{0.17}Fe_{0.08}$ HEA system and (b) Rietveld refinement profile pattern of all the peaks well fitted with C14 type hexagonal parameters with unit cell parameters a= b =5.0158 Å, c=8.1790 Å, α = β = 90°, γ = 120° under Space group $P6_3/mmc$.

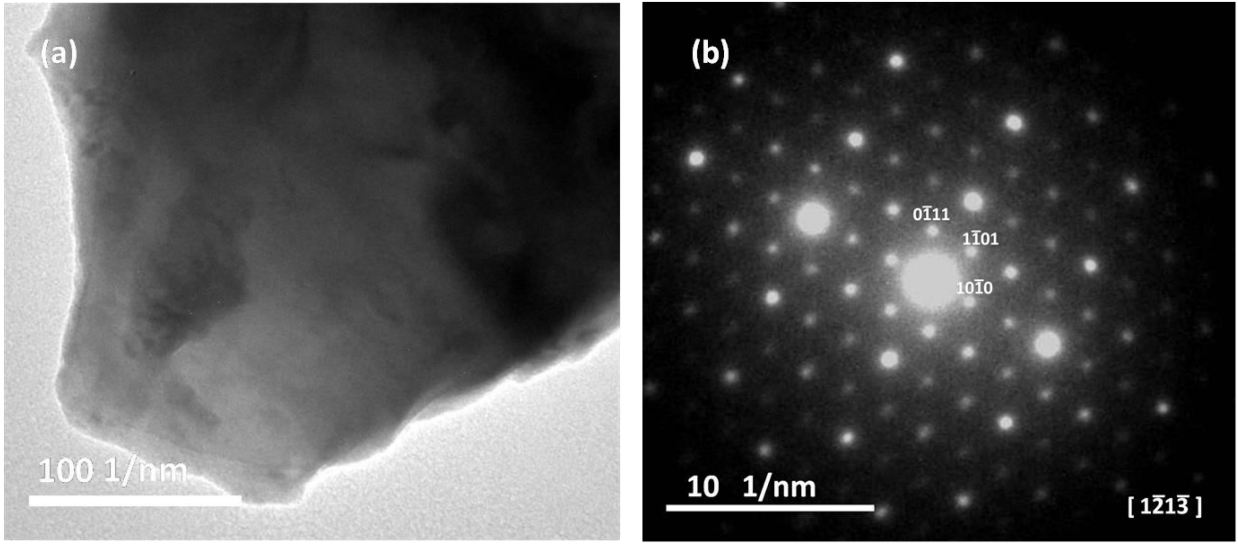

**Figure 3:** (a) TEM bright field micrograph of as-cast HEA synthesized by RF induction melting (b) Corresponding SAD pattern are shown indexed with hexagonal structure parameter under the space group of P6$_3$/mmc.

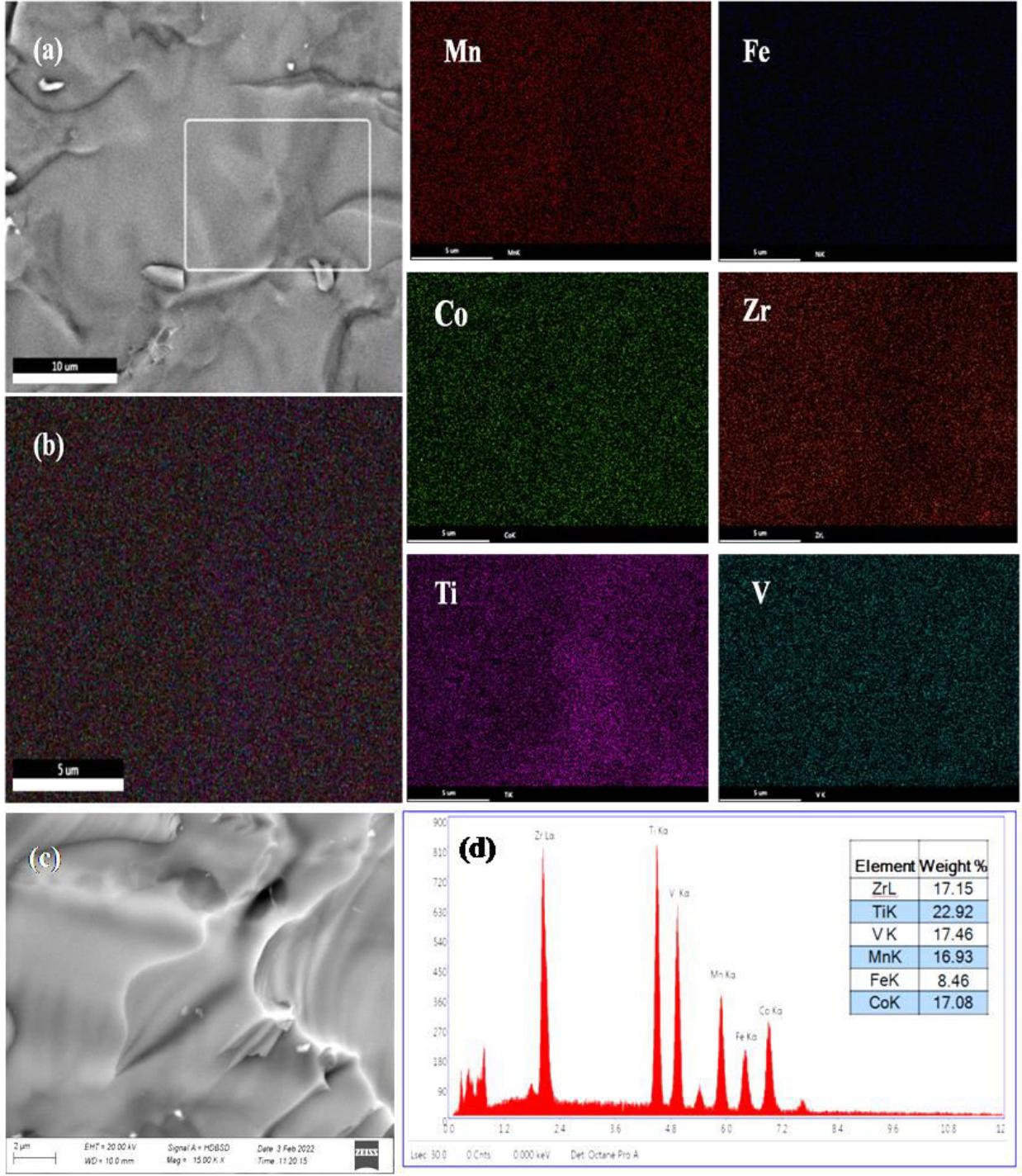

**Figure 4 :** (a) shows the SEM–BSE and energy dispersive X-ray (EDX) analysis mapping images of as cast $Ti_{0.24}V_{0.17}Zr_{0.17}Mn_{0.17}Co_{0.17}Fe_{0.08}$ HEA (b) overlays all the constituent elements present in this HEA. (c) Shows the SEM-BSE image from another region for the HEA. (d) Shows the EDX elemental spectra to validate the atomic presence of the elements in this HEA.

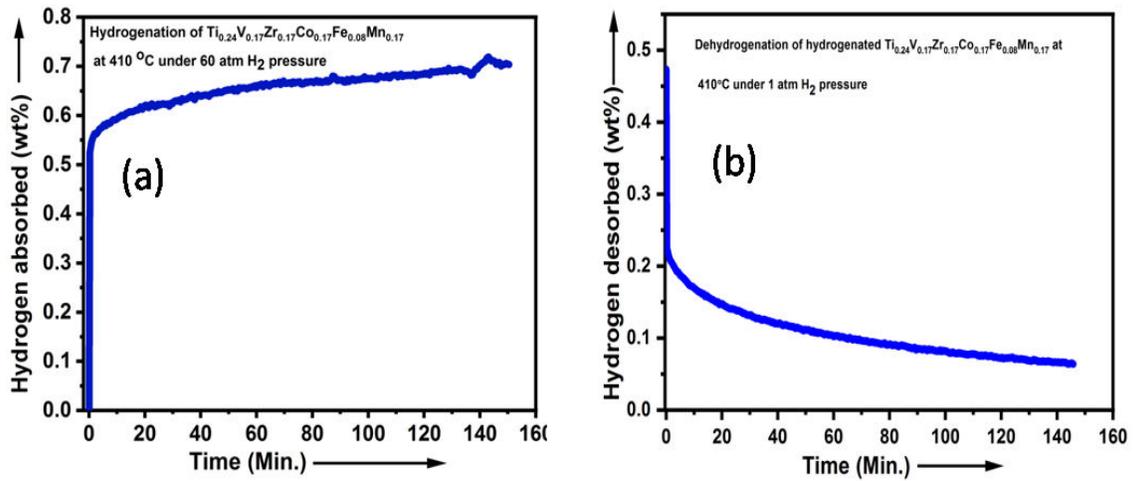

**Figure 5 :** (a) Hydrogenation curve of $Ti_{0.24}V_{0.17}Zr_{0.17}Mn_{0.17}Co_{0.17}Fe_{0.08}$ HEA at 410 ˚C under 60 atm $H_2$ pressure and (b) Dehydrogenation curve of hydrogenated $Ti_{0.24}$-$V_{0.17}$-$Zr_{0.17}$-$Co_{0.17}$-$Fe_{0.08}$-$Mn_{0.17}$ HEA at 410 ˚C under 60 atm $H_2$ pressure.

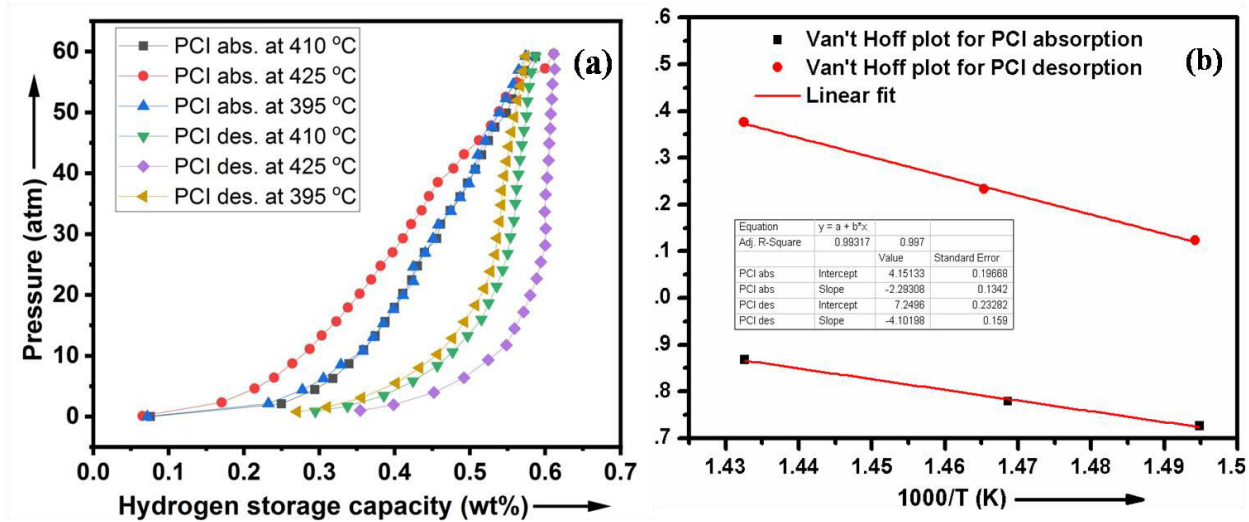

**Figure 6:**(a) Fig: (a) PCI ab/de-sorption curves of $Ti_{0.24}V_{0.17}Zr_{0.17}Mn_{0.17}Co_{0.17}Fe_{0.08}$ HEA and (b) Corresponding Van't Hoff plots for PCI ab/de-sorption curves.
16